# Applications of Intelligent Systems in Green Technology

**Sahil Mishra, Sanjaya Kumar Panda**

*Department of Computer Science and Engineering, Indian Institute of Information Technology, Design and Manufacturing, Kurnool, Andhra Pradesh*

✉ *sahilmishra32@gmail.com*

✉ *sanjayauce@gmail.com*

**ABSTRACT**

Intelligent Systems (ISs) are technologically advanced machines, which perceive and respond to the environment around them. They are usually of various forms ranging from software to hardware. ISs are generally the fusion of Artificial Intelligence (AI), robotics and Internet of things (IoT). In order to strengthen ISs, one of the key technologies is green technology (GT). It refers to the continuously advancing methods and materials, which cover techniques for producing energy to non-toxic cleaning products. It may also be broadened to saving energy, and reducing toxic and waste materials in the environment. The motto of GT can be achieved by using the ISs. In this paper, we present various applications of ISs in GT. Moreover, we discuss various possible solutions using ISs in order to overcome the on-going real-life problems.

## INTRODUCTION

Intelligent Systems (ISs) and Green Technology (GT) are the two-most emerging technologies in the field of information and communications technology. These technologies have been expanding their horizon. For instance, earlier, we just used to have simple feature phones, which could just make calls and receive calls. But now, we can do a lot of things using smart phones, which is beyond our expectations. ISs are increasingly entering our lives and pose the ability to learn from the experiences (i.e., historical data) and respond accordingly. In general, ISs consists of AI, robotics and IoT [1-3]. While AI helps a system to understand and react to certain circumstances, being just a firmware, it cannot do a lot of things [3]. Therefore, robotics does its job by taking necessary physical actions to respond to those circumstances. However, the problem is the communication among various devices constituting the system. Here, IoT plays an important role to smoothen the process of communication between the devices [2]. When these three are combined, they constitute ISs. However, IS is strengthened using one of the associated technologies, namely GT. It is the use of technology to protect the earth and the environment. It can be done in various ways. But, the most important issue to protect the environment is getting rid of all sorts of pollution. If it is not possible completely, then we have to reduce it to a tolerable limit. Similarly, deforestation needs to be fixed on urgent basis. We can use ISs to solve the above issues. There are various ways to achieve our goal.

### Pollution, Deforestation and their Causes

Earth is not what it was a million years ago. With the advancement of technology, we have been benefited a lot, but simultaneously, it has caused a lot of adverse effects on our planet's unique environment. Industrialization has led to a lot of pollutants in the air, water and soil. Vehicles are contributing as the major sources of air pollutants [3]. Even the area of agriculture is making a huge contribution of pollutants in all the three forms. Moreover, the burning of fossil fuels, mining, use of pesticides, fertilizers and use of plastic are also major pollutants. We need to fix it if we want to keep this earth habitable.

Deforestation means cutting of trees at an alarming rate [4]. It has led to a loss of habitat of flora and fauna, increase in greenhouse gases, erosion of soil and a lot more adverse effects. It is mainly caused by natural causes, like floods, parasites, etc. and human activities, such as agricultural expansion, cattle breeding, dam construction, infrastructure development, mining, oil extraction and timber extraction.

Both pollution and deforestation together have caused the rate of temperature to double in the last 50 years [5]. In this century, the computer models predict that the average temperature of earth will increase between 1.8° and 4.0° Celsius [6]. Many countries have faced the hottest days in their respective histories in 2019. Other models suggest that we will reach 7° Celsius above pre-industrial levels by 2020 [7]. We can use the ISs in various ways to control the types of pollution and deforestation up to some extent.

### Controlling Air Pollution using Intelligent Systems

Many metropolitan cities in various countries have started facing the adverse effects of air pollution, like Beijing, New Delhi etc. These cities are some of the most polluted cities of the world. Their main cause of pollution is basically vehicles and industries [3]. In order to control the pollution, the following ways may be adopted.

Firstly, if we can predict the severity of air pollution in different regions in advance, then we can shut down all the factories and restrict the usage of vehicles during that period of time. This system can be implemented using deep learning models, like Recurrent Neural Networks (RNN) [8]. The input vector data may comprise of various features, like vehicles on weekdays, vehicles on weekends, factories working hours, temperature, pressure, crop season timings, amount of gases in the atmosphere to name a few. The reason to use RNN is that we have sequential data about each day/hour.

Another change that can be done is increasing the usage of electric vehicles or vehicles driven by some renewable energy. If not possible, then the vehicles can be equipped with a system, which can detect at what speed the level of pollutants emitted by the vehicles may increase, before it actually increases. In order to implement it, the best model may be any ensemble model, which tries to improve the performance by combining the various models. The ensemble model may consist of Model A (say) to check whether all the gear mechanisms are intact and properly working. Model A can be implemented by using Gated Recurrent Unit (GRU) over Convolutional Neural Network (CNN), so that snapshots can be taken at different moments of time from a video and it can be trained using the sequence of images to check how the gear system behaves over time. This system may be used in factories to check for any fault, which may occur in the machines in the future. This can also prevent energy loss due to faulty parts or parts which may fail in the future. Another model, say B may be used to check the air pressure in tires and respective pollutant emission. On the other hand, Model C may be used to check the density of lubricant and any dirt present in it, thus predicting emission of pollutants.

ISs can be implemented as an application of AI on traffic signal system to get rid of traffic jams. Most of the times it happens that the traffic light is red on one side with a lot of traffic and it is green on the other side, where traffic is very less. It is also a cause of pollution as the vehicles' engines were on at that time, which unnecessarily releases a lot of pollutants in the air. Therefore, object detection along with localization can be used to check the density of the vehicles and give priority to that road, where density is more. It can also be helpful in giving way to emergency vehicles, like ambulances, fire-fighting van etc. Moreover, ISs can be implemented in such a manner that the engine of the vehicle automatically turns off, when it is not moving for some duration of time.

Road accidents are also a cause of traffic jams, which leads to air pollution. Therefore, a camera can be put in various accident-prone roads, which can detect accidents automatically and inform the nearby medical help team. This can be implemented using CNN on snaps of videos.

**Solving Water Related Issues using Intelligent Systems**

*Controlling Water Pollution*

The oceans are now at dangerous levels with cargo transport, offshore drilling and trash. For instance, one garbage truck of plastic is put into the oceans every one minute, which results eight million metric tons of plastic annually [9]. To remove garbage from the ocean, we need to put some drones over the ocean to detect garbage floating on its surface. The autonomous floating garbage trucks can be deployed in the areas of ocean, where there is a lot of garbage [10]. This garbage truck pulls the floating garbage, mainly plastic, by the floating rope towards a collector, which is usually a container ship. Further, this waste is recycled.

Machine Learning (ML) can be used to detect the garbage flowing into the ocean from the various places and thus knowing what sources are present in those places, which release such waste products. It can also help in determining the actions and behaviours of people in those places. This may help in preventing the waste from those sources and getting drained into water bodies. Further, it can be recycled.

Groundwater contamination can be checked using its chemical composition. But, it's not easy to extract ground water at various places and check its composition easily as it changes with time. Therefore, we can develop a system to find chemical composition using temperature, electrical conductivity and potential of hydrogen (pH) levels as input vectors. Any ML or Deep Learning (DL) model can be suitable for this task.

Coastal area's ground water quality also needs to be checked and predicted in future for usage, because the salinity of the sea also pollutes the ground water near it. We need to implement RNN to predict the future contamination as the pollutants are added and cleaned regularly. For this, we require a time-based sequential data. The input vector may comprise of location of pumping stations, previous salinity, dissolved mineral concentrations, temperature and chlorophyll levels.

River water pollution also needs to be checked as most of the cities in North India get water supply from rivers. Water quality variables, like pH, total dissolved solids, chemical oxygen demand, etc. can be taken as input vectors and the prediction of dissolved oxygen and biological oxygen demand can be made using them. The Artificial Neural Network (ANN) model can be used to make predictions. Moreover, the presence of bacteria can be done using the same model.

*Minimizing Water Wastage*

ML US alone wastes 7 billion gallons of potable water per day, which is about 11000 swimming pools [11]. We can install sensors in houses to check for daily water usage and wastage. It will strike an alarm if the water is being wasted. The amount of waste water will be determined based on the previous usage, number of people living in the houses and their daily usage. For this, we use Long-Short Term Memory (LSTM) model or GRU, which uses the data collected by the sensor.

*Preventing Ground Water Wastage*

Chennai has almost run out of ground water and it will completely run out by 2020 [12]. Despite being near coastal areas, this metropolitan city faces water issues every year. On the other hand, it gets flooded in the monsoon. The main reason behind this is that most of the city roads and footpaths are made up of concrete, which prevents the rainwater to go into the ground. The same situation is with Hyderabad, which will run out of groundwater very soon.

There are various solutions to these problems. The ground water level usually falls in summer, especially when the demand is high. Therefore, we can build a predictor to predict when the supply is going to be more. At that time, we can get alternate sources to supply the affected areas. This will reduce the load on the groundwater supply system and prevent the chaos caused due to water shortage. The input vector for the predictive model may be the pumping rate of the affected areas. We can also install rainwater syringe [13] in the areas, which may lose groundwater in the future.

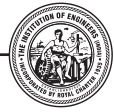

*Controlling Soil Pollution using Intelligent Systems*

The main sources of soil pollution are chemicals, leaching of wastes from landfills, oil and fuel dumping, pesticides or direct discharge of industrial wastes. To curb soil pollution, immediate measures need to be taken failing which flora is going to die very soon. Firstly, we can deploy a device to check the soil composition timely and predict when it becomes polluted. Here, we can also know that where do the pollutants come and what period of time the amount of pollutants is more. For this, we can implement RNN model as the input is time-based. This system can also be used to track waste leaking from landfills.

The main cause of soil pollution in agricultural lands is the uncontrolled usage of pesticides by the farmers. This is not only damaging the soil, but also poisons the crops and make them unfit to be consumed. Therefore, we can develop a model to suggest them the specific pesticides along with quantity that are required by the crops based on the images of the crops. The best model in this case is CNN, which will take images of crops as input to predict the diseases, and output the quantity and name of the pesticides required. The same case is also applicable to fertilizers, but they cannot be predicted just by using images of crops. We need to get the soil composition too. It is not easier to get the composition of soil quickly. As a result, we can develop a system to find chemical composition using temperature, electrical conductivity and pH levels as input vectors. This data can be easily obtained using sensors and can be used to predict the soil composition along with the name and quantity of fertilizers required.

## SUSTAINABLE LIFE USING INTELLIGENT SYSTEMS

Sustainable living is the lifestyle, which focuses on reducing the usage of earth's natural resources at the individual level and social level. The basic goal of it is to reduce carbon foot printing. There are a lot of ways to achieve it. First and the foremost thing is to predict the demand and supply of resources around the globe. It has been seen many times that this demand and supply miscommunication or broken chain has led to economic chaos. For instance, tomato crop demand and supply chain. This crop is affected by weather conditions. On the other hand, heat waves ripe it before time. Thus, farmers are forced to sell it at lower prices due to unavailability of cold storage, while consumers pay a high price for it. The areas where this crop has produced more, it is sold at very lesser prices and thus farmers get almost zero profit. On the contrary, it reaches very late after getting to know the amount of shortage, which is a long time after being bought from the farmers. Therefore, farmers sometimes destroy their crops and they don't get enough profit. This is seen more as a national economic loss, but it is more loss of earth's natural resources.

Rice, sugarcane and tomato crops require lot of water. If we destroy them, then it means the wastage of potable water, which was used. It also leads to the wastage of earth's natural resources. Moreover, a lot of fertilizers would have been used to harvest the crops, which would have again degraded the soil quality. Therefore, if we develop some system to predict demand and supply of various things across the globe and arrange the proper transportation facilities for them, then it can minimize loss of energy and make it useful in other places.

Sustainable development is not only affected by crop price inflation, but also by disease outbreaks. Many of the diseases are often concentrated among the poorest populations in the world. The disease outbreaks generally result in loss of life at a large scale, which pollutes various natural resources. If the dead bodies are not disposed off properly, then soil, air and water can get contaminated. It will also affect various parts of the world.

Using historical data, if we can predict the time, period and region of outbreaks in advance, then we can save a lot of natural resources. This prediction system still difficult to implement as it needs a lot of data to be trained. Fortunately, world health organization tries to prevent such outbreaks and reduce epidemics. Still, a lot of epidemics is occurred in various parts of the world and building a system to predict such epidemics in advance may save the lives of hundreds of people.

## CONCLUSION

In this paper, we have discussed two key technologies, namely ISs and GT. As green earth is the need of the hour, ISs can be applied to the field of GT. Moreover, as the time is passing, the earth is slowly becoming inhabitable. Therefore, people should understand that their irresponsibility is causing harm to the environment. AI can be used in making our earth a better place to live for our future generations. We can use it in various ways like to curb all types of pollution (i.e., air, water and soil). AI tools like ML and DL can be used to predict the severity of pollution in advance. They can also be used to check the source of pollution from various sources. They can even be used to predict the behaviour of people towards environment based on their activities. With the help of robotics, AI gets a helping hand to clean the environment and IoT can make various system communications, thus forming a group of ISs. These ISs can play a major role in helping to reverse the effects of human activity on the environment. They can even suggest us do's and don'ts in future to prevent any chances of future catastrophe.